\documentclass[11pt]{article}

\usepackage[latin1]{inputenc}

\usepackage[T1]{fontenc}
\usepackage[dvips]{graphicx}
\usepackage{times}

\usepackage{pstricks}
\usepackage{subeqn}

\usepackage{pifont}
\usepackage{oldgerm}
\usepackage{color} 
\usepackage{amssymb,amsbsy}
\usepackage{epic,eepic}
\usepackage{texdraw}
\usepackage{pxfonts}
\usepackage{latexsym}
\usepackage{accents}

\newtheorem{The}{Theorem}
\newtheorem{Pro}{Proposition}

\newcommand\cE{{\mathcal E}}

\title{Second order integrability conditions for difference equations. An integrable equation}
\author{Alexandre V. Mikhailov and Pavlos Xenitidis\\
School of Mathematics, University of Leeds, LS2 9JT, Leeds, UK}
\date{\today}

\begin{document}

\maketitle
\begin{abstract}
Integrability conditions for difference equations admitting a second order formal recursion operator are presented and the derivation of symmetries and canonical conservation laws is discussed. In the generic case, nonlocal conservation laws are also generated. A new integrable equation satisfying the second order integrability conditions is presented and its integrability is established by the construction of symmetries, conservation laws and a $3 \times 3$ Lax representation. Finally, the relation of the symmetries of this equation to a {\emph{generalized}} Bogoyavlensky lattice and a new integrable lattice are derived.
\end{abstract}

\section{Introduction}

The purpose of this paper is to derive some easily verifiable conditions which are necessary for the integrability of a given difference equation. With the term integrability of a difference equation, we understand the existence of an infinite hierarchy of symmetries. The derivations of such integrability conditions are based on the existence of a formal recursion operator, the general theoretical framework for which was developed in \cite{MWX}. The integrability conditions for equations admitting a first order recursion operator were given in \cite{MWX} and we refer to these conditions as first order integrability conditions. In fact, there exists a plethora of equations satisfying these first order integrability conditions \cite{GY,LY,MWX}.

However, there do exist equations which are integrable but do not admit a first order formal recursion operator. One such equation is
\begin{equation} \label{test-eq}
(u_{00}+u_{11}) u_{10} u_{01}\,+\,1\,=\,0\,,
\end{equation}
which, as far as we are aware, is new. Another equation of this type is a discrete analog of the Tzitzeica equation proposed recently by Adler \cite{A}
\begin{equation} \label{A-Tz}
u_{00} u_{11} \left(c^{-1} u_{10} u_{01}-u_{10}-u_{01}\right)\,+\,u_{11}\,+\,u_{00}\,-\,c\,=0\,.
\end{equation}
In particular, both of these equations are integrable and satisfy integrability conditions following from the existence of a second order formal recursion operator.

The main result of this paper is the derivation of these integrability conditions related to a second order formal recursion operator. We refer to these conditions as second order integrability conditions. In contrast to the case of first order integrability conditions, not all of the derived second order conditions have the form of conservation law. This is a consequence of the fact that, in general, we cannot find a fractional power of a formal difference operator with local coefficients \cite{MWX}. As we prove, if an equation admits a second order formal recursion operator but does not satisfy the first order integrability conditions, then the square root of the recursion operator cannot be computed in terms of local functions of dynamical variables and some of the integrability conditions yield nonlocal conservation laws. This differs from the continuous case where one can always compute explicitly a fractional power of any formal pseudo-differential operator \cite{Z}. Moreover, the second order 
integrability conditions provide us the means to compute the symmetries of the equation under consideration. 

All the above theoretical results are applied to equation (\ref{test-eq}) and we present its symmetries and the resulting conservation laws. We also present a Lax pair for this equation, which is given in terms of $3\times 3$ matrices, and discuss its relation to equation (\ref{A-Tz}). Finally, an interesting feature of the symmetries of equation (\ref{test-eq}) is their connection to a hierarchy of a {\emph{generalized}} Bogoyavlensky lattice 
\begin{equation} \label{gen-Bog}
\partial_{t^1} v_{0}\,=\,\left(v_0 + v_0^2\right)\,\left(v_2 v_1 \,-\,v_{-1} v_{-2}\right)
\end{equation}
via a Miura transformation. They can also be brought to the homogeneous polynomial form
\begin{equation} \label{new-pol}
\partial_{t^1} \phi_0\,=\,\phi_2\,\left(\phi_0-\phi_{-1}\right)\,+\,\phi_1\phi_{-1}\,+\,\phi_{-2}\,\left(\phi_0-\phi_1\right)-\phi_0^2
\end{equation}
by a B{\"a}cklund transformation.

The paper is organized as follows. In the first part of the next section our notation is introduced and some necessary definitions are given. The rest of Section \ref{not-theor} includes the main theoretical results on second order integrability conditions (Propositions \ref{Prop-int-cds} and \ref{Prop-int-cds-T}) and the derivation of conservation laws and symmetries. Section \ref{int-eq} is devoted mainly to the integrability aspects of equation (\ref{test-eq}) while its relation to equation (\ref{A-Tz}) is also explained. Section \ref{sym-Bog-rel} contains the integrability aspects of the symmetries for equation (\ref{test-eq}) as differential-difference equations and how they can be brought to polynomial forms (\ref{gen-Bog}) and (\ref{new-pol}) by a Miura and a B{\"a}cklund transformation, respectively. The concluding section contains an overall evaluation of the results, along with various perspectives on the subject.

\section{Formal recursion operator of order two and integrability conditions} \label{not-theor}

In this section we introduce our notation and give all the necessary definitions avoiding many technical details which can be found in \cite{MWX}. We discuss second order integrability conditions for quadrilateral equations and derive a nonlocal conservation law from these conditions.

\subsection{Notation and definitions}

In this paper, we consider scalar quadrilateral equations
\begin{equation} \label{gen-eq}
Q\left(u_{00},u_{10},u_{01},u_{11}\right)\,=\,0,
\end{equation}
where $Q$ is an irreducible polynomial depending explicitly on all of its arguments. Irreducibility means that polynomial $Q$ cannot be factorized in polynomials of lower order. Explicit dependence on the values of $u$ means that $Q_{u_{ij}} \ne 0$, where 
\begin{equation}\label{Q-der-def}
Q_{u_{ij}} \,\coloneqq\, \frac{\partial Q}{\partial u_{ij}}.
\end{equation}
The shift operators in the $n$ and the $m$ direction are denoted respectively by $\cal{S}$ and $\cal{T}$ and their action can be defined as
$${\cal{S}}^p{\cal{T}}^q\,:\,u\,\mapsto\,u_{pq} \,\equiv\,u(n+p,m+q)\,.$$

In the theory of difference equations, $u_{pq}$ are treated as variables and we denote the set of all shifts of variable $u$ by $U = \{u_{pq} | (p,q) \in {\mathbb{Z}}^2 \}$. In the case when $Q$ in equation (\ref{gen-eq}) is a multi-linear polynomial, one can eliminate uniquely all variables $u_{pq}$, with $pq \ne 0$, using the equation and its shifts. More precisely, we can solve equation (\ref{gen-eq}) for any value of $u$, i.e.
\begin{equation}
 \label{subsQ}
\begin{array}{ll}
u_{0,0}=F(u_{1,0},u_{0,1},u_{1,1}),\qquad &u_{1,0}=G(u_{0,0},u_{0,1},u_{1,1}), \\
u_{0,1}=H(u_{0,0},u_{1,0},u_{1,1}),\qquad &u_{1,1}=M(u_{0,0},u_{1,0},u_{0,1}),
\end{array}
\end{equation}
where $F,G,H$ and $M$ are rational functions of their arguments. Now we can define recursively the elimination map $\cal{E}$, \cite{MWX}, as
\begin{equation}\label{Emap}
\begin{array}{ll}
\forall p\in {\mathbb{Z}},\ \qquad &{\cal{E}}(u_{0,p})=u_{0,p},\qquad  \cE(u_{p,0})=u_{p,0}\, ,\\
{\rm if}\ p>0,q>0,\  \quad &{\cal{E}}(u_{p,q})=M(\cE(u_{p-1,q-1}),\cE(u_{p,q-1}),\cE(u_{p-1,q}))\,,\\
{\rm if}\ p<0,q>0,\  \quad &\cE(u_{p,q})=H(\cE(u_{p,q-1}),\cE(u_{p+1,q-1}),\cE(u_{p+1,q}))\,,\\
{\rm if}\ p>0,q<0,\  \quad &\cE(u_{p,q})=G(\cE(u_{p-1,q}),\cE(u_{p-1,q+1}),\cE(u_{p,q+1}))\,,\\
 {\rm if}\ p<0,q<0,\ \quad &\cE(u_{p,q})=F(\cE(u_{p+1,q}),\cE(u_{p,q+1}),\cE(u_{p+1,q+1}))\, .
 \end{array}
\end{equation}
Hence, any expression in variables $U$ can be reduced to an expression involving only variables $u_{\ell 0}$ and $u_{0\ell}$. We refer to these variables as dynamical variables, and we denote their corresponding sets by
$$U_{\bf{s}} \,=\, \left\{ u_{\ell 0}\left| \ell \in {\mathbb{Z}} \right. \right\}\,,\quad U_{\bf{t}} \,=\, \left\{ u_{0\ell}\left| \ell \in {\mathbb{Z}} \right. \right\}\,,\quad U_0 = U_{\bf{s}} \cup U_{\bf{t}}\,.$$
Moreover, from now on, we denote the field of rational functions of the dynamical variables as
$${\cal{F}}_{\bf{s}} \,=\,{\mathbb{C}}\left(U_{\bf{s}}\right)\,,\quad {\cal{F}}_{\bf{t}} \,=\,{\mathbb{C}}\left(U_{\bf{t}}\right)\,,\quad {\cal{F}}_{\bf{0}} \,=\,{\mathbb{C}}\left(U_{\bf{0}}\right),$$
respectively.

A symmetry $K$ of equation (\ref{gen-eq}) is an element of ${\cal{F}}_0$ satisfying the equation
\begin{equation}\label{det-eq}
{\cal{E}}\left(\sum_{p,q} Q_{u_{pq}} {\cal{S}}^p{\cal{T}}^q\left(K\right)\right)\,=\,0\,. 
\end{equation}
For quadrilateral equations (\ref{gen-eq}), symmetries can always be written as a sum of an element of ${\cal{F}}_{\bf{s}}$ and element from ${\cal{F}}_{\bf{t}}$ \cite{TTX}. In what follows we consider symmetries involving only dynamical variables $U_{\bf{s}}$ or variables $U_{\bf{t}}$. An {\bf{s}}-pseudo-difference operator $\mathfrak{R}$ is a recursion operator of equation (\ref{gen-eq}) if it maps any symmetry $K \in {\cal{F}}_{\bf s}$ of (\ref{gen-eq}) to another symmetry of the same equation \cite{MWX}. Similarly, a {\bf{t}}-pseudo-difference operator $\hat{\mathfrak{R}}$ is a recursion operator of equation (\ref{gen-eq}) if it maps any symmetry $\hat{K} \in {\cal{F}}_{\bf t}$ of (\ref{gen-eq}) to another symmetry of the same equation. In our derivations, pseudo-difference operators are represented by their formal Laurent series \cite{MWX}. Specifically, if $\mathfrak{R}$ is an $N$th order {\bf{s}}-pseudo-difference operator, then its formal series will be denoted by ${\mathfrak{R}}_{\rm L}$ and we will 
write
$${\mathfrak{R}}_{\rm L}\,=\,r_N {\cal{S}}^N + r_{N-1} {\cal{S}}^{N-1} + \cdots + r_0 + r_{-1} {\cal{S}}^{-1} + \cdots,\quad r_i \in {\cal{F}}_{\bf {s}}\,.$$
In the same way, an $N$th order {\bf{t}}-pseudo-difference operator $\hat{\mathfrak{R}}$ will be represented by
$$\hat{\mathfrak{R}}_{\rm L}\,=\,\hat{r}_N {\cal{T}}^N + \hat{r}_{N-1} {\cal{T}}^{N-1} + \cdots + \hat{r}_0 + \hat{r}_{-1} {\cal{T}}^{-1} + \cdots,\quad \hat{r}_i \in {\cal{F}}_{\bf {t}}\,.$$

We define a {\emph{local conservation law}} for equation (\ref{gen-eq}) as a pair of functions $\rho, \sigma \in {\mbox{Span}}_{\mathbb{C}} ({\cal{F}}_0 , \log {\cal{F}}_0)$ such that
$${\cal{E}}\Big(\left({\cal{T}}-{\bf{1}}\right)\left(\rho\right)\,-\,\left({\cal{S}}-{\bf{1}}\right)\left(\sigma\right)\Big)\,=\,0\,.$$
We refer to functions $\rho$ and $\sigma$ as density and flux, respectively, of the local conservation law. A conservation law is trivial if $(\rho,\sigma)$ is the gradient of some element $h \in  {\mbox{Span}}_{\mathbb{C}} ({\cal{F}}_0 , \log {\cal{F}}_0)$, i.e. $\rho= \left({\cal{S}}-{\bf{1}}\right)(h)$ and $\sigma = \left({\cal{T}}-{\bf{1}}\right)(h)$.

For densities depending only on dynamical variables $U_{\bf s}$, we define equivalence classes \cite{MWX,MWX2}. Since the field $\cal{F}_{\bf s}$ is a linear space over $\mathbb{C}$,  we consider the extended linear space ${\cal{L}}_{\bf s} = {\mbox{Span}}_{\mathbb{C}} ({\cal{F}}_{\bf s} , \log {\cal{F}}_{\bf s})$. Hence, elements of ${\cal{L}}_{\bf s}$ are linear combinations with complex coefficients of elements in $\cal{F}_{\bf s}$ and logarithms of elements in $\cal{F}_{\bf s}$. For any element $f \in {\cal L}_{\bf s}$ we define an equivalence class (or a {\emph{functional}}) by saying that two elements $f, g \in {\cal L}_{\bf s}$ are equivalent if  $f-g = \left({\cal{S}}-{\bf 1}\right)\left(h\right)$ for some $h \in {\cal{L}}_{\bf s}$. The space of functionals will be denoted by ${\cal{F}}^\prime_{\bf s}$, it is a linear space over $\mathbb{C}$ and it does not inherit a ring or field structure of $\cal{F}_{\bf s}$.

The last definition we have to recall from \cite{MWX} is the order of a conservation law. For the conservation laws we discuss in the next sections, the densities are elements of ${\cal{F}}^\prime_{\bf s}$ for which  the order can be defined invariantly, i.e. independently from the representative of the equivalence class. Let us first define the order of an element of ${\cal{F}}_{\bf{s}}$. If $f(u_{k 0},\ldots,u_{ \ell 0}) \in {\cal{F}}_{\bf{s}}$, where $k < \ell$ and  $f_{u_{k 0}} f_{u_{ \ell 0}} \ne 0$, then the order of $f$ is defined as
$${\rm{ord}} f \,\coloneqq \,(k,\ell)\,.$$
The variational derivative of $f \in {\cal{F}}^\prime$ is defined as 
$$\delta_{\bf{s}}\left(f\right) \,\coloneqq\,\sum_{j=k}^{\ell} {\cal{S}}^{-j}\left(\frac{\partial f}{\partial u_{j0}}\right)\,, $$
respectively. If $\rho \in {\cal{F}}^\prime_{\bf s}$ is a density of a conservation law, then its order ${\rm{ord}}_{\bf{s}}\left(\rho\right) $ is defined as 
$${\rm{ord}}_{\bf{s}}\left(\rho\right) \,:=\,N_2 - N_1,\quad {\mbox{where}} \quad (N_1,N_2) = {\rm{ord}} \left(\delta_{\bf{s}}(\rho) \right)\,.$$

\subsection{Integrability conditions}

In our derivations for integrability conditions \cite{MWX}, we expand the pseudo-difference operator 
\begin{equation} \label{F-op}
\Phi\,=\,\left(Q_{u_{11}} {\cal{S}} + Q_{u_{01}}\right)^{-1}\circ\left(Q_{u_{10}} {\cal{S}} + Q_{u_{00}}\right) 
\end{equation}
and its inverse $\Phi^{-1}$ in formal Laurent series,
\begin{equation}\label{FL-op}
\Phi_{\rm{L}}\,=\,\alpha_0\,+\,\alpha_{1} {\cal{S}}^{-1} \,+\,\alpha_{2} {\cal{S}}^{-2} \,+\,\cdots,\qquad \Phi_{\rm{L}}^{-1}\,=\,\beta_0\,+\,\beta_{1} {\cal{S}}^{-1} \,+\,\beta_{2} {\cal{S}}^{-2} \,+\,\cdots
\end{equation}
where the coefficients $\alpha_i$, $\beta_i$ have the following form.
\begin{equation}\label{a-b-cf}
\begin{array}{lcl}
\alpha_{0} = {\cal{S}}^{-1}\left(\frac{Q_{u_{10}}}{Q_{u_{11}}} \right) &\qquad& \beta_{0} = {\cal{S}}^{-1}\left(\frac{Q_{u_{11}}}{Q_{u_{10}}} \right) \\
\alpha_{1} = {\cal{S}}^{-1}\left(\frac{Q_{u_{00}}}{Q_{u_{11}}}\,-\,\frac{Q_{u_{01}}}{Q_{u_{11}}}{\cal{S}}^{-1} \left(\frac{Q_{u_{10}}}{Q_{u_{11}}} \right)\right) && \beta_{1} = {\cal{S}}^{-1}\left(\frac{Q_{u_{01}}}{Q_{u_{10}}}\,-\,\frac{Q_{u_{00}}}{Q_{u_{10}}}{\cal{S}}^{-1}\left(\frac{Q_{u_{11}}}{Q_{u_{10}}}\right) \right)\\
\alpha_{k+1} =(-1)^{k} {\cal{S}}^{-1}\left(\frac{Q_{u_{01}}}{Q_{u_{11}}} \alpha_{k} \right) && \beta_{k+1} = (-1)^{k} {\cal{S}}^{-1}\left(\frac{Q_{u_{00}}}{Q_{u_{10}}} \beta_{k} \right),\quad k \ge 1
\end{array} 
\end{equation}

In \cite{MWX}, we proved the following proposition which gives the integrability conditions for a scalar quadrilateral equation admitting a first order recursion operator.
\begin{Pro}[First order integrability conditions \cite{MWX}] \label{Prop-1-ic}
If equation $Q(u_{00},u_{10},u_{01},u_{11})= 0$ admits a first order formal recursion operator
$${\mathfrak{Q}}_{\rm{L}}\,=\,q_1 {\cal{S}}+q_0+q_{-1}{\cal{S}}^{-1}+\cdots\,,\quad q_i \in {\cal{F}}_{\bf{s}}\,,$$
then the following integrability conditions must hold
\begin{subequations}\label{cond0}
\begin{eqnarray}
&& ({\cal{T}}-{\bf 1})(\log q_1)=({\cal{S}}-{\bf 1}){\cal{S}}^{-1}\left(\log\frac{Q_{u_{11}}}{Q_{u_{10}}}\right),\label{condp1}\\
&& ({\cal{T}}-{\bf 1})( q_0)=({\cal{S}}-{\bf 1}){\cal{S}}^{-1}(q_1 F),\label{condp0} \\
&& ({\cal{T}}-{\bf 1})(q_{-1}{\cal{S}}^{-1}(q_1)+q_0^2+q_1{\cal{S}}(q_{-1}))=({\cal{S}}-{\bf 1})(\sigma_2), \label{condpt}
\end{eqnarray} 
\end{subequations}
where
\[
\sigma_2 = {\cal{S}}^{-1}(q_1\, F) \left\{ {\cal{S}}^{-1}(q_0)+ q_0 - {\cal{S}}^{-2}\left(q_1  F\right)\right\}-(1+{\cal{S}}^{-1}) \left( q_1  G {\cal{S}}^{-1} \left(q_1 F\right) \right),
\]
and $F,G$ denote
\[
 F=\frac{Q_{u_{01}}}{Q_{u_{10}}} {\cal{S}}^{-1}\left(\frac{Q_{u_{10}}}{Q_{u_{11}}}\right)- \frac{Q_{u_{00}}}{Q_{u_{10}}},\qquad
G=\frac{Q_{u_{00}}}{Q_{u_{10}}}.
\]
\end{Pro}

The first theoretical result of this paper is the derivation of integrability conditions for an equation admitting a second order recursion operator. The proof of this proposition is omitted here because it is similar to the proof of Proposition \ref{Prop-1-ic} in \cite{MWX}.
\begin{Pro}[Second order integrability conditions] \label{Prop-int-cds}
If equation $Q(u_{00},u_{10},u_{01},u_{11})= 0$ admits a second order formal recursion operator
\begin{equation}
{\mathfrak{R}}_{\rm{L}}\,=\,r_2 {\cal{S}}^2\,+\,r_1 {\cal{S}} \,+\,r_0\,+\,r_{-1} {\cal{S}}^{-1}\,+\,r_{-2} {\cal{S}}^{-2}\,+\,\cdots\,, \quad r_i\,\in\,{\cal{F}}_{\bf{s}}\,,
\end{equation}
then the following integrability conditions must hold
\begin{subequations} \label{int-cond}
\begin{eqnarray}
&&\left({\cal{T}}-{\bf{1}}\right)\left(\log  r_2\right)\,=\,\left({\cal{S}}^{2}-{\bf{1}}\right) {\cal{S}}^{-1} \left(\log \frac{Q_{u_{11}}}{Q_{u_{10}}}\right)\,,\label{int-cond-1}\\
&&{\cal{T}}\left(r_1\right)\,-\,\alpha_0\,r_1\,{\cal{S}}\left(\beta_0\right)\,=\,\alpha_0 r_2 {\cal{S}}^2\left(\beta_1\right)\,+\,\alpha_1 {\cal{S}}^{-1} \left(r_2\right) {\cal{S}}\left( \beta_0\right)\,,\label{int-cond-3}\\
&&\left({\cal{T}}-{\bf{1}}\right)\,\left(r_0\right)\,=\,\left({\cal{S}}-{\bf{1}}\right){\cal{S}}^{-1}\,\left\{\left(\alpha_0 r_1 + \alpha_1 {\cal{S}}^{-1} \left( r_2\right)\right) {\cal{S}}\left( \beta_1\right) + \alpha_0 r_2 {\cal{S}}^2\left( \beta_2\right) + {\cal{S}}^{-1}\left(\alpha_0 r_2 \right) {\cal{S}}\left(\beta_2\right) \right\}\,, \label{int-cond-2}
\end{eqnarray} 
\end{subequations}
where functions $\alpha_i$, $\beta_i$ are given in (\ref{a-b-cf}).
\end{Pro}
Here we restricted ourselves to present only the first three conditions. One could compute more conditions since there are no obstacles in our construction but the resulting expressions are becoming longer and more complicated. It is very likely that if an equation satisfies these three conditions then it should be integrable.

It is apparent from conditions (\ref{int-cond}) that only the even coefficients of the corresponding recursion operators provide us with canonical conservation laws of equation (\ref{gen-eq}), whereas in the case of first order recursion operator all the integrability conditions have the form of conservation laws \cite{MWX}. If we can find a first order formal series $\mathfrak{Q}$ such that ${\mathfrak{Q}}^2 = {\mathfrak{R}}_{\rm{L}}$, then the residues of the odd powers of $\mathfrak{Q}$ will give us the missing conservation laws. Although in the differential case one can always compute explicitly the fractional power ${\mathfrak{R}}^{1/N}_{\rm{L}}$ of a formal series of a pseudo-differential operator of order $N \ge 2$, \cite{Z}, this is not always possible in the difference case \cite{MW, MWX}. For the difference case, we prove the following
\begin{The} \label{Thr-nl-cl}
Suppose that equation $Q(u_{00},u_{10},u_{01},u_{11})= 0$ does not satisfy the first order integrability condition (\ref{condp1}) and admits a second order formal recursion operator ${\mathfrak{R}}_{\rm{L}}=r_2 {\cal{S}}^2 + r_1 {\cal{S}} + r_0 + \cdots$, $r_i \in {\cal{F}}_{\bf{s}}$. Then
\begin{enumerate}
\item There exists no first order formal series ${\mathfrak{Q}}= q_1 {\cal{S}} + q_0 + q_{-1} {\cal{S}}^{-1}+ \cdots$, with $q_i \in {\cal{F}}_{\bf{s}}$, such that ${\mathfrak{Q}}^2 = {\mathfrak{R}}_{\rm{L}}$.
\item The pair of functions
\begin{equation}\label{nonloc-cl}
\rho_1\,=\,\frac{r_1}{w}\,,\qquad \sigma_1\,=\,\left({\cal{S}}+{\bf{1}}\right){\cal{S}}^{-1}\left[w\,\left(\frac{Q_{u_{01}}}{Q_{u_{10}}} {\cal{S}}^{-1}\left(\frac{Q_{u_{10}}}{Q_{u_{11}}}\right) \,-\,\frac{Q_{u_{00}}}{Q_{u_{10}}}\right)\right]\,,
\end{equation}
where $w$ is an additional variable (a potential) such that
\begin{equation} \label{c1-eq}
{\cal{S}}\left(w\right)\,=\,\frac{r_2}{w}\,,\quad {\cal{T}}\left(w\right)\,=\,\frac{Q_{u_{11}}}{Q_{u_{10}}} {\cal{S}}^{-1}\left(\frac{Q_{u_{10}}}{Q_{u_{11}}}\right)\,w\,,
\end{equation}
defines a {\rm{nonlocal}} conservation law for equation $Q(u_{00},u_{10},u_{01},u_{11})= 0$.
\end{enumerate}
\end{The}
\noindent {\bf{Proof}} Suppose that there exists a first order formal series $\mathfrak{Q}$ with coefficients in ${\cal{F}}_{\bf{s}}$ such that ${\mathfrak{Q}}^2 = {\mathfrak{R}}_{\rm{L}}$. Then $q_1 {\cal{S}}(q_1) = r_2$ and  condition (\ref{int-cond-1}) can be written as
$$\left({\cal{T}}-{\bf{1}}\right)\left(\log  q_1 \right)\,=\,\left({\cal{S}}-{\bf{1}}\right) {\cal{S}}^{-1} \left(\log \frac{Q_{u_{11}}}{Q_{u_{10}}}\right)\,.$$
This is nothing else but integrability condition (\ref{condp1}) and it contradicts our hypothesis. Hence, coefficients of $\mathfrak{Q}$ cannot be in ${\cal{F}}_{\bf{s}}$. 

It can be easily verified that system (\ref{c1-eq}) is consistent since its compatibility condition
$$\left\{\frac{{\cal{T}}(r_2)}{r_2}\,-\,{\cal{S}}\left(\frac{Q_{u_{11}}}{Q_{u_{10}}}\right) {\cal{S}}^{-1}\left(\frac{Q_{u_{10}}}{Q_{u_{11}}}\right) \right\}\,\frac{r_2}{w}\,\frac{Q_{u_{10}}}{Q_{u_{11}}}\,{\cal{S}}^{-1}\left(\frac{Q_{u_{11}}}{Q_{u_{10}}}\right)\,=\,0$$
holds in view of condition (\ref{int-cond-1}). Moreover, using  system (\ref{c1-eq}) to eliminate the shifts $w_{i0}$, $w_{0i}$ from $\left({\cal{T}}-1\right)\left(\rho_1\right) - \left({\cal{S}}-1\right)\left(\sigma_1\right)$, and subsequently applying the elimination map $\cal{E}$, the resulting expression is equivalent to condition (\ref{int-cond-3}) and hence equal to zero. \hfill $\Box$

The first coefficients of the formal recursion operator help us to compute the symmetries of a given equation. If we denote by $K^{(i)}$ the symmetries of the equation, then the first three symmetries satisfy the following relations.
\begin{equation} \label{r-K-conne}
\begin{array}{l}
\partial_{u_{20}}K^{(1)}= r_2\,,\quad \partial_{u_{10}}K^{(1)} = r_1\,, \\
\partial_{u_{40}}K^{(2)}= r_2 {\cal{S}}^2\left(r_2\right)\,,\quad \partial_{u_{30}}K^{(2)}= r_2 {\cal{S}}^2\left(r_1\right) + r_1 {\cal{S}}\left(r_2\right)\,,\quad \partial_{u_{20}}K^{(2)}=r_2 {\cal{S}}^2\left(r_0\right) + r_1 {\cal{S}}\left(r_1\right) + r_0 r_2\,,\\
\partial_{u_{60}}K^{(3)}= r_2 {\cal{S}}^2\left(r_2\right){\cal{S}}^4\left(r_2\right)\,,\quad \partial_{u_{50}}K^{(3)} = r_2 {\cal{S}}^2\left(r_2\right){\cal{S}}^4\left(r_1\right) \,+\,\left(r_2 {\cal{S}}^2\left(r_1\right) + r_1 {\cal{S}}\left(r_2\right)\right){\cal{S}}^3\left(r_2\right)\,,\\
\partial_{u_{40}}K^{(3)} = r_2 {\cal{S}}^2\left(r_2\right){\cal{S}}^4\left(r_0\right)\,+\,\left(r_2 {\cal{S}}^2\left(r_1\right) + r_1 {\cal{S}}\left(r_2\right)\right){\cal{S}}^3\left(r_1\right) + \left(r_2 {\cal{S}}^2\left(r_0\right) + r_1 {\cal{S}}\left(r_1\right) + r_0 r_2\right) {\cal{S}}^2\left(r_2\right)\,.
\end{array} 
\end{equation}
These relations determine partially the forms of those symmetries, e.g. the first two relations determine $K^{(1)}$ up to an arbitrary function of $u_{00}$, $u_{-10}$ and $u_{-20}$. The arbitrary functions then can be determined from condition (\ref{det-eq}).

So far we considered only the $n$ direction of the lattice. Similar results hold for formal recursion operators, integrability conditions, symmetries and conservation laws for the other lattice direction. We present only the relevant results here without any proofs. In that case, operator $\Phi$ is replaced by 
\begin{equation} \label{Ps-op}
\Psi\,=\,\left(Q_{u_{11}} {\cal{T}} + Q_{u_{10}}\right)^{-1}\circ\left(Q_{u_{01}} {\cal{T}} + Q_{u_{00}}\right) \,,
\end{equation}
and for the corresponding formal Laurent series of $\Psi$ and its inverse $\Psi^{-1}$,
\begin{equation}\label{PL-op}
\Psi_{\rm{L}}\,=\,a_0\,+\,a_{1} {\cal{T}}^{-1} \,+\,a_{2} {\cal{T}}^{-2} \,+\,\cdots,\qquad \Psi_{\rm{L}}^{-1}\,=\,b_0\,+\,b_{1} {\cal{S}}^{-1} \,+\,b_{2} {\cal{S}}^{-2} \,+\,\cdots\,,
\end{equation}
one can find that
\begin{equation}\label{ab-cf}
\begin{array}{lcl}
a_{0} = {\cal{T}}^{-1}\left(\frac{Q_{u_{01}}}{Q_{u_{11}}} \right) &\qquad& b_{0} = {\cal{T}}^{-1}\left(\frac{Q_{u_{11}}}{Q_{u_{01}}} \right) \\
a_{1} = {\cal{T}}^{-1}\left(\frac{Q_{u_{00}}}{Q_{u_{11}}}\,-\,\frac{Q_{u_{10}}}{Q_{u_{11}}}{\cal{T}}^{-1} \left(\frac{Q_{u_{01}}}{Q_{u_{11}}} \right)\right) && b_{1} = {\cal{T}}^{-1}\left(\frac{Q_{u_{10}}}{Q_{u_{01}}}\,-\,\frac{Q_{u_{00}}}{Q_{u_{01}}}{\cal{T}}^{-1}\left(\frac{Q_{u_{11}}}{Q_{u_{01}}}\right) \right)\\
a_{k+1} =(-1)^{k} {\cal{T}}^{-1}\left(\frac{Q_{u_{10}}}{Q_{u_{11}}} a_{k} \right) && b_{k+1} = (-1)^{k} {\cal{T}}^{-1}\left(\frac{Q_{u_{00}}}{Q_{u_{01}}} b_{k} \right),\quad k \ge 1
\end{array} 
\end{equation}

\begin{Pro}[First order integrability conditions \cite{MWX}] \label{Prop-2-ic}
If equation $Q(u_{00},u_{10},u_{01},u_{11})= 0$ admits a first order formal recursion operator
$$\hat{{\mathfrak{Q}}}_{\rm{L}}\,=\,\hat{q}_1 {\cal{T}}+\hat{q}_0+\hat{q}_{-1}{\cal{T}}^{-1}+\cdots\,,\quad \hat{q}_i \in {\cal{F}}_{\bf{t}}\,,$$
then the following integrability conditions must hold
\begin{subequations}\label{condm0}
\begin{eqnarray}
&& ({\cal{S}}-{\bf 1})(\log \hat{q}_1)=({\cal{T}}-{\bf 1}){\cal{T}}^{-1}\left(\ln\frac{Q_{u_{11}}}{Q_{u_{01}}}\right),\label{condmp1}\\
&& ({\cal{S}}-{\bf 1})( \hat{q}_0)=({\cal{T}}-{\bf 1}){\cal{T}}^{-1}(\hat{q}_1 \hat{F}),\label{condmp0} \\
&& ({\cal{S}}-{\bf 1})(\hat{q}_{-1}{\cal{T}}^{-1}(\hat{q}_1)+\hat{q}_0^2+\hat{q}_1{\cal{T}}(\hat{q}_{-1}))=({\cal{T}}-{\bf 1})(\hat{\sigma}_2), \label{condmpt}
\end{eqnarray} 
\end{subequations}
where
\[
\hat{\sigma}_2 = {\cal{T}}^{-1}(\hat{q}_1\, \hat{F}) \left\{ {\cal{T}}^{-1}(\hat{q}_0)+ \hat{q}_0 - {\cal{T}}^{-2}\left(\hat{q}_1 \hat{F}\right)\right\}-(1+{\cal{T}}^{-1}) \left(\hat{q}_1 \hat{G} {\cal{T}}^{-1} \left(\hat{q}_1 \hat{F}\right) \right),
\]
and $\hat{F},\hat{G}$ denote
\[
\hat{F}=\frac{Q_{u_{10}}}{Q_{u_{01}}} {\cal{T}}^{-1}\left(\frac{Q_{u_{01}}}{Q_{u_{11}}}\right)- \frac{Q_{u_{00}}}{Q_{u_{01}}},\qquad
\hat{G}=\frac{Q_{u_{00}}}{Q_{u_{01}}}.
\]
\end{Pro}

\begin{Pro}[Second order integrability conditions] \label{Prop-int-cds-T}
If equation (\ref{gen-eq}) admits a second order formal recursion operator
\begin{equation}
\hat{\mathfrak{R}}_{\rm{L}}\,=\,\hat{r}_2 {\cal{T}}^2\,+\,\hat{r}_1 {\cal{T}} \,+\,\hat{r}_0\,+\,\hat{r}_{-1} {\cal{T}}^{-1}\,+\,\hat{r}_{-2} {\cal{T}}^{-2}\,+\,\cdots\,,\quad \hat{r}_i \in {\cal{F}}_{\bf{t}}\,,
\end{equation}
then the following integrability conditions must hold
\begin{subequations} \label{int-cond-T}
\begin{eqnarray}
&&\left({\cal{S}}-{\bf{1}}\right)\left(\log \hat{r}_2\right)\,=\,\left({\cal{T}}^{2}-{\bf{1}}\right) {\cal{T}}^{-1} \left(\log \frac{Q_{u_{11}}}{Q_{u_{01}}}\right)\,,\\
&&{\cal{S}}\left(\hat{r}_1\right)\,-\,a_0\,\hat{r}_1\,{\cal{T}}\left(b_0\right)\,=\,a_0 \hat{r}_2 {\cal{T}}^2\left(b_1\right)\,+\,a_1 {\cal{T}}^{-1} \left(\hat{r}_2\right) {\cal{T}}\left( b_0\right)\,,\\
&&\left({\cal{S}}-{\bf{1}}\right)\,\left(\hat{r}_0\right)\,=\,\left({\cal{T}}-{\bf{1}}\right){\cal{T}}^{-1}\,\left\{\left(a_0 \hat{r}_1 + a_1 {\cal{T}}^{-1} \left(\hat{r}_2\right)\right) {\cal{T}}\left(b_1\right) + a_0 \hat{r}_2 {\cal{T}}^2\left(b_2\right) + {\cal{T}}^{-1}\left(a_0 \hat{r}_2 \right) {\cal{T}}\left(b_2\right) \right\}\,,
\end{eqnarray} 
\end{subequations}
where functions $a_i$, $b_i$ are given in (\ref{ab-cf}).
\end{Pro}

\begin{The} \label{Thr-nl-cl-T}
Suppose that equation $Q(u_{00},u_{10},u_{01},u_{11})= 0$ does not satisfy the first order integrability condition (\ref{condmp1}) and admits a second order formal recursion operator $\hat{\mathfrak{R}}_{\rm{L}}=\hat{r}_2 {\cal{T}}^2 + \hat{r}_1 {\cal{T}} + \hat{r}_0 + \cdots$, $\hat{r}_i \in {\cal{F}}_{\bf{t}}$. Then
\begin{enumerate}
\item There exists no first order formal series $\hat{\mathfrak{Q}}= \hat{q}_1 {\cal{T}} + \hat{q}_0 + \hat{q}_{-1} {\cal{T}}^{-1}+ \cdots$, with $\hat{q}_i \in {\cal{F}}_{\bf{t}}$, such that $\hat{\mathfrak{Q}}^2 = \hat{\mathfrak{R}}_{\rm{L}}$.
\item The pair of functions
\begin{equation}\label{nonloc-cl-T}
\hat{\rho}_1\,=\,\frac{\hat{r}_1}{\hat{w}}\,,\qquad \hat{\sigma}_1\,=\,\left({\cal{T}}+1\right){\cal{T}}^{-1}\left[\hat{w}\,\left(\frac{Q_{u_{10}}}{Q_{u_{01}}} {\cal{T}}^{-1}\left(\frac{Q_{u_{01}}}{Q_{u_{11}}}\right) \,-\,\frac{Q_{u_{00}}}{Q_{u_{01}}}\right)\right]\,,
\end{equation} 
where $\hat{w}$ is an additional variable (a potential) such that
\begin{equation} \label{c1-eq-T}
{\cal{T}}\left(\hat{w}\right)\,=\,\frac{\hat{r}_2}{\hat{w}}\,,\quad {\cal{S}}\left(\hat{w}\right)\,=\,\frac{Q_{u_{11}}}{Q_{u_{01}}} {\cal{T}}^{-1}\left(\frac{Q_{u_{01}}}{Q_{u_{11}}}\right)\,\hat{w}\,.
\end{equation}
defines a {\rm{nonlocal}} conservation law for equation $Q(u_{00},u_{10},u_{01},u_{11})= 0$.
\end{enumerate}
\end{The}

\section{A new integrable equation} \label{int-eq}

In this section we study the properties of equation
\begin{equation} \label{q-eq}
(u_{00}+u_{11}) u_{10} u_{01}\,+\,1\,=\,0\,,
\end{equation}
which can be summarized in the following list.
\begin{enumerate}
\item Equation (\ref{q-eq}) is not linearizable by a point transformation.
\item It does not admit any first integrals, i.e. {\bf{s}}- and {\bf{t}}-constants in the terminology of \cite{MWX}.
\item It is integrable in the sense that
\begin{enumerate}
\item it satisfies the second order integrability conditions (\ref{int-cond}), (\ref{int-cond-T}) and
\item it admits a Lax pair given in terms of $3 \times 3$ matrices.
\end{enumerate}
\item It can be viewed as a degeneration of the Tzitzeica equation proposed recently by Adler \cite{A}.
\end{enumerate}
Because equation (\ref{q-eq}) is invariant under the interchange of lattice directions, $u_{ij} \mapsto u_{ji}$, we present here only the results deriving from Proposition \ref{Prop-int-cds} and Theorem \ref{Thr-nl-cl}, i.e. the $n$ lattice direction. Corresponding symmetries and conservation laws for the other lattice direction follow from the presented results by simply interchanging indices and shifts operators.

\subsection{Point symmetries and lower order conservation laws}

We start by presenting the point symmetries and the lower order conservation laws of equation (\ref{q-eq}). In particular, equation (\ref{q-eq}) admits two points symmetries generated by
\begin{equation} \label{sym-lie}
\frac{\partial u_{00}}{\partial \epsilon_k}\,=\, \chi^{k (n-m)} u_{00}\,,\quad k=1,2,
\end{equation}
where $\chi$ is a primitive cube root of unity $\chi^2 + \chi + 1 = 0$. Its second-order conservation laws are given by the pair $(f_\omega, \sigma_\omega)$, where
\begin{equation} \label{cl-3p}
f_\omega\,=\,\omega^{n-m}\,\left(\omega\,u_{00}u_{10}\,-\,\frac{1}{u_{10}}\right)\,,\quad g_{\omega}\,=\,\omega^{n-m}\left(u_{00} u_{01}\,-\,\frac{\omega}{u_{01}}\right)\,,\quad \omega^3\,=\,1\,.
\end{equation}

\subsection{Integrability conditions, higher order conservation laws and generalized symmetries}

It can be easily verified that there does not exist any function $q_1 \in {\cal{F}}_{\bf{s}}$ such that condition (\ref{condp1}) holds for equation (\ref{q-eq}). Hence, this equation does not satisfy the first order integrability conditions (\ref{cond0}). But, it does satisfy the second order integrability conditions (\ref{int-cond}).

Indeed, using integrability conditions (\ref{int-cond}), one can compute the first three coefficients of the formal recursion operator. More precisely, introducing the function 
\begin{equation} \label{def-F-G}
F_{00} := u_{10} u_{00} u_{-10} - 1\,,
\end{equation}
those coefficients can be written as follows.
\begin{subequations}\label{r2-r0}
\begin{eqnarray}
&& r_2\,=\,-\,\frac{u_{10} u_{00}^2}{F_{10}^2 F_{00}}\,,\label{r2} \\
&& r_1\,=\,-\,\frac{u_{00}}{u_{10}}\,\left[\frac{1}{F_{10}F_{00}}\left(1 + \frac{1}{F_{00}}\right)\,+\,\left({\cal{S}}-{\bf{1}}\right)\left\{\frac{1}{F_{00}F_{-10}}\left(1 + \frac{1}{F_{00}}\right)\right\}  \right]\,,\label{r1} \\
&& r_0 \,=\,-\,\frac{3}{F_{10} F_{00}}\,-\,\frac{3}{F_{10} F_{00} F_{-10}}\,+\,\left({\cal{S}}-{\bf{1}}\right)\left\{\frac{1}{F_{-10}F_{-20}}\left(1+\frac{1}{F_{00}}\right) - \,\frac{1}{F_{00}F_{-10}}\left(\frac{1}{F_{00}}+\frac{1}{F_{-10}}\right) \right\}\,.\label{r0}
\end{eqnarray}
\end{subequations}
These functions and relations (\ref{int-cond-1}), (\ref{int-cond-2}) provide us the two first members of the hierarchy of local canonical conservation laws admitted by equation (\ref{q-eq}), which are equivalent to
\begin{subequations} \label{cl-high}
\begin{eqnarray}
&& \rho_{0}\,=\,\log \frac{F_{00}}{u_{00}}\,,\qquad\qquad\qquad\quad \sigma_{0}\,=\,\log \left(u_{-10}+ u_{01}\right)\,,\label{cl-high-1}\\
&& \rho_{2}\,=\,\frac{1}{F_{10} F_{00}}\,+\,\frac{1}{F_{10} F_{00} F_{-10}}\,,\quad \sigma_{2}\,=\,\frac{-1}{F_{00} F_{-10}}\,.\label{cl-high-2}
\end{eqnarray}
\end{subequations}
It can be easily verified that ${\rm{ord}}_{\bf{s}}\left(\rho_0\right) = 4$ and ${\rm{ord}}_{\bf{s}}\left(\rho_2\right) = 8$. Moreover, using Theorem \ref{Thr-nl-cl}, we introduce the potential $w$ through the relations
\begin{equation}\label{c1-q}
w\,{\cal{S}}\left(w\right)\,=\,-\,\frac{u_{10} u_{00}^2}{F_{10}^2 F_{00}}\,,\quad {\cal{T}}\left(w\right)\,=\,-\,\frac{u_{10}^2 u_{01} (u_{-10}+u_{01})}{u_{00}}\,w\,,
\end{equation}
to construct the nonlocal conservation law for equation (\ref{q-eq})
\begin{subequations}\label{nl-cl-q}
\begin{eqnarray}
&&\rho_1\,=\,\frac{-1}{w}\,\frac{u_{00}}{u_{10}}\,\left[\frac{1}{F_{10}F_{00}}\left(1 + \frac{1}{F_{00}}\right)\,+\,\left({\cal{S}}-{\bf{1}}\right)\left\{\frac{1}{F_{00}F_{-10}}\left(1 + \frac{1}{F_{00}}\right)\right\}  \right]\,,\\
&& \sigma_1\,=\,\left({\cal{S}}+{\bf{1}}\right){\cal{S}}^{-1}\left(\frac{u_{10} (u_{00} u_{10} u_{01}^2 +u_{01}+u_{-10})}{u_{00} u_{01}}\,w\right)\,.
\end{eqnarray} 
\end{subequations}

Equation (\ref{q-eq}) admits a hierarchy of generalized symmetries which can be found using relations (\ref{r-K-conne}) and the expressions (\ref{r2-r0}) for the leading coefficients of the recursion operator. The first two symmetries are given by
\begin{subequations} \label{sym-gen}
\begin{eqnarray}
&&\frac{\partial u_{00}}{\partial t^1} = K^{(1)} := u_{00} ({\cal{S}}-{\bf{1}}) \frac{1}{F_{00} F_{-10}}\,,\\
&& \frac{\partial u_{00}}{\partial t^{2}}= K^{(2)} := u_{00} ({\cal{S}}-{\bf{1}}) \left[\frac{-1}{F_{00} F_{-10}}\left\{\frac{1}{F_{00} F_{-10}} + \left({\cal{S}}+{\bf{1}}\right) \left(1+\frac{1}{F_{-10}}\right)\left(\frac{1}{F_{10} F_{00}}+\frac{1}{F_{-20} F_{-30}} \right)\right\}\right]\,.
\end{eqnarray}
\end{subequations}
The  expression for the third symmetry is too involved to present it here. In the next section we show that after difference substitutions, equations (\ref{sym-gen}) and the next symmetry of this hierarchy can be presented in an elegant polynomial form.

\subsection{Lax representation}

The linear system
\begin{equation} \label{lax-pair}
\Psi_{10}=\left(\begin{array}{ccc} 0 & 1 & 0 \\ -u_{00} & -u_{00} u_{10} & \lambda \\-1 &0&\frac{1}{u_{00}}\end{array} \right) \Psi_{00},\quad  \Psi_{01}=\left(\begin{array}{ccc} 0 & 0 & 1 \\-1 &0&\frac{1}{u_{00}}\\ \frac{u_{00}u_{01}}{\lambda} & \frac{-1}{\lambda} & 0 \end{array} \right) \Psi_{00}
\end{equation}
provides a Lax pair for equation (\ref{q-eq}). This system is consistent if and only if equation (\ref{q-eq}) holds.

\subsection{Degeneration of Tzitzeica equation}

The discrete analog of Tzitzeica equation \cite{A}
\begin{equation} \label{Adler-Tz}
u_{00} u_{11} \left(c^{-1} u_{10} u_{01}-u_{10}-u_{01}\right)\,+\,u_{11}\,+\,u_{00}\,-\,c\,=0\,.
\end{equation}
satisfies all conditions of Propositions \ref{Prop-int-cds} and \ref{Prop-int-cds-T} and does not satisfy the first order integrability conditions. Its two first symmetries and local conserved densities can be found in \cite{A}. If we use relations (\ref{r-K-conne}) to compute $r_2$, $r_1$ and $r_0$ from the symmetries of equation (\ref{Adler-Tz}), then Proposition \ref{Prop-int-cds} and Theorem \ref{Thr-nl-cl} provides us with two local conserved densities along with their corresponding fluxes and a nonlocal conservation law for (\ref{Adler-Tz}).

Equation (\ref{q-eq}) can be viewed as a degeneration of equation (\ref{Adler-Tz}). To be more precise, if we set
$$u_{ij}\,\rightarrow\,\frac{\epsilon}{u_{ij}}\,,\qquad c \,\rightarrow \,\epsilon^3,$$
into (\ref{Adler-Tz}), then it becomes
\begin{equation} \label{e-Adler-Tz}
\epsilon \left(\frac{1}{u_{00}}\,+\,\frac{1}{u_{11}}\,+\,\frac{1}{u_{00}u_{10}u_{01}u_{11}} \right) \,-\,\epsilon^3 \left( 1\,+\,\frac{1}{u_{00} u_{10} u_{11}}\,+\,\frac{1}{u_{00} u_{01} u_{11}} \right)\,=\,0.
\end{equation}
Multiplying the above equation by $\epsilon^{-1}$ and taking the limit $\epsilon \rightarrow 0$, equation (\ref{q-eq}) follows. If we divide (\ref{e-Adler-Tz}) by $\epsilon^2$ and take the limit $\epsilon \rightarrow \infty$, we arrive at
$$ u_{00} u_{10} u_{01} u_{11} + u_{10}+u_{01}\,=\,0\,.$$
This equation can be mapped to (\ref{q-eq}) by the transformation $u_{ij} \mapsto 1/u_{-i,j}$ .

\section{Hierarchy of integrable differential -- difference equations} \label{sym-Bog-rel}

In this section we focus on the differential-difference equations (\ref{sym-gen}), and discuss their integrability. We also present a Miura transformation which brings this hierarchy to a polynomial form which generalizes Bogoyavlensky lattices and differ from the lattices studied in \cite{AP}. Finally, we discuss a potentiation of these systems and derive a new hierarchy of differential-difference equations which is given in terms of homogeneous polynomials.

Since in our considerations in this section variable $m$ does not vary, we simplify our notation omitting the second index from $u$ and $F$ and  rewrite equations (\ref{sym-gen}) as
\begin{subequations} \label{dif-dif}
\begin{eqnarray}
&&\frac{\partial u_{0}}{\partial t^1} \,=\, u_{0} ({\cal{S}}-{\bf{1}}) \frac{1}{F_{0} F_{-1}}\,,\label{dd-1}\\
&&\frac{\partial u_{0}}{\partial t^{2}} \,=\, u_{0} ({\cal{S}}-{\bf{1}}) \left[\frac{-1}{F_{0} F_{-1}}\left\{\frac{1}{F_{0} F_{-1}} + \left({\cal{S}}+{\bf{1}}\right) \left(1+\frac{1}{F_{-1}}\right)\left(\frac{1}{F_{1} F_{0}}+\frac{1}{F_{-2} F_{-3}} \right)\right\}\right]\,, \label{dd-2}
\end{eqnarray} 
\end{subequations}
where
\begin{equation} \label{dd-F}
F_0 \,:=\,u_{1}u_{0}u_{-1}\,-\,1\,. 
\end{equation}

\subsection{Integrability aspects}

The point symmetries of equations (\ref{dif-dif}) are generated by
\begin{equation}\label{dd-p-sym}
\frac{\partial u_0}{\partial \epsilon_k}\,=\,\chi^{k\,n} u_0,\quad k=1,2,\qquad \chi^2+\chi+1\,=\,0\,.
\end{equation}

A Lax representation for equations (\ref{dif-dif}) can be systematically derived starting with the $n$ part of system (\ref{lax-pair}). In particular, for equation (\ref{dd-1}) the Lax pair is given by the following linear system.
\begin{subequations}\label{dd-Lax}
\begin{eqnarray}
&&\Psi_{1}=\left(\begin{array}{ccc} 0 & 1 & 0 \\ -u_{0} & -u_{0} u_{1} & \lambda \\-1 &0&\frac{1}{u_{0}}\end{array}\right)\Psi_0\\
&&\frac{\partial \Psi_0}{\partial t^1} =\left\{\frac{1}{\lambda-1} \left(\begin{array}{ccc} -\frac{1}{3} -\frac{1}{F_{0}} \left(1+\frac{1}{F_1}\right) & \frac{-u_1}{F_0} \left(1+\frac{1}{F_1}\right) & \frac{1}{u_0 F_0}\left(1+ \frac{1}{F_1}\right) \\ \frac{1}{u_1}\left(1+ \frac{1}{F_0}\right) & \frac{2}{3} + \frac{1}{F_0} & \frac{-1}{u_0 u_1}\left(1+ \frac{1}{F_0}\right) \\ \frac{-u_0}{F_0F_1} & \frac{-u_0 u_1}{F_0 F_1} & - \frac{1}{3} + \frac{1}{F_1 F_0}  \end{array} \right)\,+\,\left(\begin{array}{ccc}0&0&\frac{u_{-1} u_{-2}}{F_0F_{-1}} \\ \frac{-u_0 u_{-1}}{F_1F_0} &0& \frac{-u_{-1}}{F_0}\\ 0&0&0 \end{array} \right) \right\}\Psi_0
\end{eqnarray} 
\end{subequations}

Apparently, equations (\ref{dif-dif}) can be written in a conserved form with density $\log u_0$. Other densities are given from the densities of conservation laws for equation (\ref{q-eq}), and in particular $f_\omega$ in (\ref{cl-3p}) and $\rho_0$, $\rho_2$ in (\ref{cl-high}). Fluxes can be easily found, and for (\ref{dd-1}) we have
\begin{subequations} \label{dd-cl}
\begin{eqnarray}
&&\partial_{t^1} \left\{\omega^{n}\,\left(\omega\,u_{0}u_{1}\,-\,\frac{1}{u_{1}}\right)\right\} \,=\,({\cal{S}}-{\bf{1}})  \left(\frac{\omega^{n-1}}{u_{0} F_{0}} \left(1 + \frac{1}{F_{1}}\right) + \frac{\omega^{n}}{u_{1} F_{1}} \left(1 + \frac{1}{F_{0}}\right) + \frac{\omega^{n+1} u_{0} u_{1}}{F_{0}F_{-1}}\right)\,,\label{dd-cl-0} \\
&& \partial_{t^1} \left(\log F_{0}\right) \,=\,\left({\cal{S}}-{\bf{1}}\right)\left[ \left({\cal{S}}+{\bf{1}}\right)\left(\frac{1}{F_{-1} F_{-2}} \left(1+ \frac{1}{F_{0}}\right)\right) \,+\,\frac{1}{F_{1} F_{0}} \right] \,,\label{dd-cl-2} \\
&& \partial_{t^1} \left(\frac{1}{F_{1} F_{0}}\,+\,\frac{1}{F_{1} F_{0} F_{-1}} \right) \,=\,-\,\left({\cal{S}}-{\bf{1}}\right)\left( \frac{(F_{-1}+1)(F_{-2}+1)}{F_{1} F_{0}^2 F_{-1}^2 F_{-2}}\,+\,\left({\cal{S}}+{\bf{1}}\right)\frac{(F_{-1}+1)(F_{-3}+1)}{F_{1} F_{0} F_{-1}^2 F_{-2} F_{-3}} \right)\,.\label{dd-cl-8}
\end{eqnarray} 
\end{subequations}
Similarly, a nonlocal conservation law follows from (\ref{nl-cl-q}) and for equation (\ref{dd-1}) can be written as
\begin{subequations} \label{dd-nl-cl}
\begin{equation}
\partial_{t^1} \left(\frac{-1}{w_0}\,\frac{u_{0}}{u_{1}}\,\left[\frac{1}{F_{1}F_{0}}\left(1 + \frac{1}{F_{0}}\right)\,+\,\left({\cal{S}}-{\bf{1}}\right)\left\{\frac{1}{F_{0}F_{-1}}\left(1 + \frac{1}{F_{0}}\right)\right\}\right]\,\right)\,=\,\left({\cal{S}}^2-{\bf{1}}\right){\cal{S}}^{-1}\left(A\,w_{0}\right)\,, 
\end{equation}
where function $A$ is given by
\begin{equation}
A \coloneqq \frac{u_1}{u_0}\,\left\{\frac{1}{F_0 F_{-1}}\left(\frac{1}{F_1}+\frac{1}{F_{-2}}\right)\,-\,\frac{1}{F_1 F_2} \left(\frac{2}{F_0}  + \frac{1}{F_{-1}}\right) \,+\,\frac{1}{F_{-1}F_{-2}}\,-\,\frac{2}{F_1 F_2} \right\}
\end{equation}
and potential $w$ is defined by the system
\begin{equation}
w_0\,w_1\,=\,-\,\frac{u_{1} u_{0}^2}{F_{1}^2 F_{0}}\,,\quad \partial_{t^1} w_0\,=\,-w_0\,\left({\cal{S}}-{\bf{1}}\right) \left\{\frac{1}{F_{-1} F_{-2}} \left(1 + \frac{1}{F_0} \right)\,+\,\frac{2}{F_{1} F_{0}} \left(1 + \frac{1}{F_{-1}} \right)\right\}\,. 
\end{equation}
\end{subequations}

\subsection{Miura transformation and Bogoyavlensky type lattices}

One can easily verify that the Miura transformation
\begin{equation} \label{Miura}
v_0 \,=\,\frac{1}{u_1 u_0 u_{-1} -1}\,. 
\end{equation}
 maps equations (\ref{dif-dif}) and the next symmetry of this hierarchy (which has been omitted in the previous section) to
\begin{equation} \label{comb-BmB}
\begin{array}{l}
\partial_{t^1} v_0 \,=\,\left(v_0^2+v_0\right) \left(v_2 v_1 - v_{-1}v_{-2} \right)\,,\\
\partial_{t^{2}} v_0 \,=\,\left(v_0^2+v_0\right)  \left( B^{(2)}\,+\,M^{(2)}\right)\,,\\
\partial_{t^{3}} v_0 \,=\,\left(v_0^2+v_0\right)  \left( B^{(3)}\,+\,M^{(3)}\,-\,P^{(3)}\right)\,.
\end{array}
\end{equation}
Polynomials $B^{(i)}$, $M^{(i)}$ are related to the Bogoyavlensky and the modified Bogoyavlensky lattices \cite{B}
\begin{equation} \label{B-mB-lattice}
\partial_{x^i} v_0\,=\, v_0\,B^{(i)}\,,\qquad\partial_{y^i} v_0\,=\,v_0^2\,M^{(i)}\,,\qquad i\,=\,1,2,\cdots,
\end{equation}
respectively. In particular, the first member of both hierarchies involve the same polynomial,
\begin{equation} \label{B1-M1-pol}
B^{(1)} = M^{(1)} = v_2 v_1 - v_{-1} v_{-2}\,,
\end{equation}
and polynomials $B^{(2)}$, $M^{(2)}$ are given in the Appendix where also $P^{(3)}$ can be found. Polynomials $B^{(3)}$, $M^{(3)}$ can be computed either from $B^{(2)}$, $M^{(2)}$ using the recursion operators for lattices (\ref{B-mB-lattice}) \cite{JPW}, or in terms of specific homogeneous polynomials as described in \cite{S}. It is obvious that the first equation in (\ref{comb-BmB}) is the sum of the first two flows $x^1$ and $y^1$. These two flows do not commute and thus the integrability of their sum is an exceptional property, see also \cite{AP} where similar observations were presented.

A Lax pair for lattices (\ref{comb-BmB}) can be derived from Lax pair (\ref{dd-Lax}) in a systematic way. Here we present explicitly this Lax pair for the first flow.
\begin{subequations} \label{LP-BmB}
\begin{eqnarray}
&&\Phi_3 = \left(\begin{array}{ccc} 1-\lambda + \frac{1}{v_0} & \frac{v_{-1}+1}{v_0 v_{-1}} & \frac{-\lambda}{v_0}\\ \frac{(\lambda-1)v_0 -1}{(v_1+1) v_0} & \frac{-((\lambda-1) v_1 v_0 +1) (v_{-1}+1)}{(v_1+1) v_0 v_{-1}} & \frac{\lambda}{(v_1+1) v_0} \\ \frac{v_1}{(v_1+1) v_0} & \frac{v_1 (v_{-1}+1)}{(v_1+1) v_0 v_{-1}} & \frac{-v_1 ((\lambda-1) v_0+\lambda)}{(v_1+1) v_0} \end{array} \right) \Phi_0\,,\\
&& \nonumber \\
&&\frac{\partial \Phi_0}{\partial t^1}  = \left(\begin{array}{ccc} \frac{3 v_{-1} (v_{-2}+1)+1}{3 (\lambda-1)} & \frac{(v_{-1}+1)(v_{-2}+1)}{\lambda-1} & \frac{-\lambda v_{-1} (v_{-2}+1)}{\lambda-1} \\ \frac{((\lambda-1) v_0-1)v_{-1}}{\lambda-1} & \frac{3 ((\lambda-1) v_0-1) v_{-1} - 3 (\lambda-1) v_{-2} v_{-3} -2}{3 (\lambda-1)} & \frac{\lambda v_{-1}}{\lambda-1}\\ \frac{v_{-1} v_{-2}}{\lambda-1} & \frac{(v_{-1}+1) v_{-2}}{\lambda-1} & v_0 v_{-1} + \frac{1-3 \lambda v_{-1} v_{-2}}{3 (\lambda-1)} \end{array} \right) \Phi_0\,.
\end{eqnarray} 
\end{subequations}

All equations (\ref{comb-BmB}) have the form of conservation law with density $\log(v_0/(v_0+1))$, and, for instance, the first flow can be written as
$$ \partial_{t^1} \log\left(\frac{v_0}{v_0+1}\right)\,=\,\left( {\cal{S}}^3-{\bf{1}}\right) \left(v_{-1} v_{-2}\right)\,.$$
Another zeroth order conserved density is $\log (v_0)$ which follows from the fourth order density in (\ref{dd-cl-2}). The corresponding conservation law for the first member of hierarchy (\ref{B-mB-lattice}) reads as follows.
$$ \partial_{t^1} \log\left(v_0\right)\,=\,\left( {\cal{S}}^2-{\bf{1}}\right) \left(v_0 v_{-1} v_{-2}\right)\,+\,\left( {\cal{S}}^3-{\bf{1}}\right) \left(v_{-1} v_{-2}\right)\,.$$
The fourth order density $\varrho_2 = v_1 v_0 (v_{-1}+1)$ follows actually from the eighth order conserved density in (\ref{dd-cl-8}).
$$\partial_{t^1} \Big(v_1 v_0 (v_{-1}+1)\Big) \,=\,\left( {\cal{S}}-{\bf{1}}\right)\Big\{v_1 v_0^2 v_{-1} (v_{-1}+1) (v_{-2}+1) + \left({\cal{S}}+{\bf{1}}\right) \left(v_1 v_0 v_{-1} (v_{-1}+1) v_{-2} (v_{-3}+1) \right)\Big\}\,.$$
It seems that under the Miura transformation, canonical conserved densities of equations (\ref{dif-dif}) are mapped to conserved densities of equations (\ref{B-mB-lattice}) but the orders of these densities are reduced by four. Finally, in a similar way, a nonlocal conservation law can be derived from (\ref{dd-nl-cl}), which for the first equation in (\ref{comb-BmB}) can be written as
$$ \partial_{t^1} \left(\frac{v_1 v_0 (v_1+v_0+1) - (v_0+1) v_0 v_{-1}}{w_0}\right)=\left({\cal{S}}- {\cal{S}}^{-1}\right) \Big(w_0\Big\{v_2 v_1 (2 v_0 +v_{-1}+2) -v_0 v_{-1} (v_1 + v_{-2}) - v_{-1} v_{-2}\Big\} \Big),$$ 
where
$$w_1 w_0 = v_1 (v_1+1) v_0,\quad \partial_{t^1} w_0=\left({\cal{S}}-{\bf{1}}\right) \Big\{\left({\cal{S}}+{\bf{1}}\right)(v_0 v_{-1}) + v_{-1} v_{-2} +v_0 v_{-1} (v_{-2}+2 v_1)\Big\} w_0. $$

\subsection{Potentiation and a new hierarchy of differential-difference equations}

Starting with equations (\ref{dif-dif}) and employing the third conserved density in (\ref{dd-cl}), one can introduce a potential $\phi$ via
\begin{equation} \label{pot-x}
\phi_1\,-\,\phi_0\,=\,\frac{1}{F_{1} F_{0}}\,+\,\frac{1}{F_{1} F_{0} F_{-1}}\,.
\end{equation}
In terms of the potential $\phi$, equations (\ref{dif-dif}) and the next symmetry of the hierarchy can be written as 
\begin{equation}\label{dd-pot}
\begin{array}{l}
\partial_{t^1} \phi_0=N^{(1)}\,\coloneqq\,\phi_2 \left(\phi_0-\phi_{-1}\right)+ \phi_1 \phi_{-1} + \phi_{-2} \left(\phi_0-v_1\right)  - \phi_0^2\,,\\
\partial_{t^2} \phi_0 = N^{(2)} \,\coloneqq\,\phi_4 (\phi_2-\phi_1) (\phi_0-\phi_{-1}) + \phi_3 (\phi_1-\phi_0) (\phi_0-\phi_{-2})\\
{\phantom{\partial_{t^2} \phi_0 = N^{(2)} \,\coloneqq\,}}\,-\,\phi_{-4} (\phi_{-2}-\phi_{-1}) (\phi_0-\phi_{1}) - \phi_{-3} (\phi_{-1}-\phi_0) (\phi_0-\phi_{2})\,,\\
 \partial_{t^3} \phi_0 = N^{(3)} \,\coloneqq\,P_{+} + P_{-} + (\phi_3 \phi_{-3} +2 \phi_2 \phi_{-2}) (\phi_1-\phi_0)(\phi_{-1}-\phi_0) + \phi_1 \phi_{-1} (\phi_1 \phi_{-1}-\phi_0^2),
\end{array}
\end{equation}
where
\begin{equation}
\begin{array}{cl}
P_{+}& = \phi_6 (\phi_4-\phi_3)(\phi_2-\phi_1) (\phi_0-\phi_{-1}) + \\
 & \phantom{+} \left\{\phi_5 (\phi_3-\phi_2) + \phi_4 (\phi_2-\phi_1) + \phi_3 ( \phi_1-\phi_0)\right\} \left\{\phi_2 (\phi_0-\phi_{-1})+\phi_1 (\phi_{-1}-\phi_{-2}) + \phi_0 (\phi_{-2}-\phi_0) \right\}+\\
& \phantom{+} \phi_4 (\phi_2-\phi_1)(\phi_{-1}-\phi_0) \phi_{-3} + \phi_2^2 (\phi_{-1}-\phi_0)^2 + \phi_2 (\phi_{-1}-\phi_0) (\phi_0^2 -2 \phi_1 \phi_{-1})\,,
\end{array} 
\end{equation}
and $P_-$ follows from $P_+$ by changing $\phi_i \rightarrow \phi_{-i}$. Alternatively, equations (\ref{dd-pot}) can be derived from (\ref{comb-BmB}). The connection among (\ref{comb-BmB}) and (\ref{dd-pot}) is given by
\begin{equation}
\phi_1\,-\,\phi_0\,=\, v_1 v_0 (v_{-1}+1)\,.
\end{equation}

As far as we are aware, equations (\ref{dd-pot}) constitute a new hierarchy of integrable lattices. The polynomials $N^{(i)}$ involved in the right hand sides of these equations are homogeneous with respect to variables $\phi_i$, and their corresponding degrees are $\deg N^{(i)} =i+1$, $i=1,2,\ldots$. Polynomials $N^{(2 k-1)}$, respectively $N^{(2 k)}$, where $k=1,2,\ldots$, are symmetric, respectively antisymmetric, under the interchange $\phi_i \leftrightarrow \phi_{-i}$.

\section{Conclusions}

In this paper we derived and presented second order integrability conditions for quadrilateral difference equations based on the framework developed in \cite{MWX}. These integrability conditions are given in Propositions \ref{Prop-int-cds} and \ref{Prop-int-cds-T}, and the nonlocal conservation laws deriving from those integrability conditions are presented in Theorems \ref{Thr-nl-cl} and \ref{Thr-nl-cl-T}.

In the case of the first order integrability conditions given in Propositions \ref{Prop-1-ic} and \ref{Prop-2-ic}, every condition has the form of local conservation law. In this way, a hierarchy of canonical conservation laws can be derived. The situation however is different with the second order integrability conditions. Only the even coefficients of the recursion operator provide us with local conservation laws. Nonlocal conservation laws can be derived from the second order integrability conditions presented here and they are given explicitly in Theorems \ref{Thr-nl-cl} and \ref{Thr-nl-cl-T}. 

The latter Theorems also explain why it is not always possible to find explicitly the fractional powers of a recursion operator in the discrete case, in contrast to the continuous case where this is always possible. In the continuous case, one always considers first order integrability conditions and it may happen that some of these conditions result to trivial conservation laws. In the discrete case, if an equation does not satisfy the first of the first order integrability conditions then it leads to a nonlocal conservation law. 

As far as we are aware, equations (\ref{q-eq}) and (\ref{Adler-Tz}) are, so far, the only examples of quadrilateral equations satisfying the second order integrability conditions and not the first order integrability conditions. A systematic classification of such equations based on the integrability conditions presented here is an open and challenging problem.

Returning to equations (\ref{q-eq}) and (\ref{Adler-Tz}), the former equation is related to the latter by a degeneration, as all the ABS equations are related to Q4 under certain degenerations \cite{ABS,AS,NAH}. Both equations admit a Lax representation with $3 \times 3$ matrices and their symmetries are related via Miura transformations to generalizations of the Bogoyavlensky lattices, see equations (\ref{B-mB-lattice}) and references \cite{A,AP}. A potentiation of symmetries (\ref{dif-dif}) leads to the hierarchy of integrable lattices (\ref{dd-pot}) which seems to be new. In fact, it can be easily verified that the equations of this hierarchy are the symmetries in the $n$ direction for the difference equation
\begin{equation} \label{6p-eq}
\left(\phi_{10} \,-\,\phi_{11}\right)\left( \phi_{00}-\phi_{01}\right) \left(\phi_{-10}-\phi_{-11}\right)\,-\,\left(\phi_{11}-\phi_{00}\right) \left(\phi_{01}-\phi_{-10}\right)\,=\,0\,.
\end{equation}
This difference equation is multi-linear, is defined on two consecutive quadrilaterals on the lattice and its relation to equation (\ref{q-eq}) is given by the potentiation
$$\phi_{10}-\phi_{00}\,=\,\frac{1}{F_{10} F_{00}}\,+\,\frac{1}{F_{10} F_{00} F_{-10}}\,,\quad \phi_{01}-\phi_{00}\,=\,\frac{-1}{F_{00} F_{-10}}\,. $$
Equation (\ref{6p-eq}) is out of the class of equations we consider in this paper but it would be interesting to extend the theory of formal recursion operators and integrability conditions to study equations of this form. In particular for equation (\ref{6p-eq}), it would be interesting to study its symmetries in the $m$ direction, as well as to derive a Lax representation for this equation and its hierarchy of symmetries (\ref{dd-pot}).

\section*{Acknowledgments}
The authors acknowledge support from the EPSRC grant {\emph{Structure of partial difference equations with continuous symmetries and conservation laws}}, EP/I038675/1.

\section*{Appendix: Bogoyavlensky lattices}

We present here for convenience the polynomials involved in the first symmetry of the Bogoyavlensky lattices (\ref{B-mB-lattice}). Polynomials $B^{(2)}$, $M^{(2)}$ can be written as
\begin{subequations} \label{B-mB-sym-pol}
\begin{equation}
B^{(2)}\,\coloneqq\,B^{(2)}_+ \,-\,B^{(2)}_-\,,\qquad M^{(2)}\,\coloneqq\,M^{(2)}_+ \,-\,M^{(2)}_-\,, 
\end{equation}
where
\begin{equation}
B^{(2)}_+ \,\coloneqq\,  v_4 v_3 v_2 v_1 + v_3 v_2^2 v_1 + v_2^2 v_1^2 + v_2 v_1^2 v_0 + v_2 v_1 v_0 v_{-1}\,,\quad
M^{(2)}_+ \coloneqq v_4 v_3 v_2^2 v_1 + v_3 v_2^2 v_1^2 + v_2^2 v_1^2 v_0 + v_2 v_1^2 v_0 v_{-1}\,,
\end{equation}
\end{subequations}
and $B^{(2)}_-$ and $M^{(2)}_-$ follow from $B^{(2)}_+$ and $M^{(2)}_+$, respectively, by changing $v_i \rightarrow v_{-i}$. Using the same notation, polynomial $P^{(3)}$ involved in the third equation of (\ref{comb-BmB}) can be written as
$$P^{(3)} \,\coloneqq\, P^{(3)}_+ \,-\,P^{(3)}_-\,,$$
where
\begin{eqnarray} \label{Pn-pol}
P^{(3)}_+ &\coloneqq& v_6 v_5 v_4 v_3 v_2 v_1 (v_4+v_2) + v_5 v_4 v_3 v_2 v_1 (v_4 v_3 + v_4 v_2 + v_3 v_2 + v_2 v_1) + 2 v_4^2 v_3^2 v_2^2 v_1+\nonumber \\
&& v_4 v_3 v_2 v_1 (2 v_3 v_2^2 + 2 v_3 v_2 v_1 + 2 v_2^2 v_1 + 3 v_2 v_1 v_0 + v_2 v_0 v_{-1} + v_1 v_0 v_{-1}) + 2 v_3^2 v_2^3 v_1^2 + \nonumber \\
&& v_3 v_2^2 v_1^2 (2 v_2 v_1 + 2 v_2 v_0 +2 v_1 v_0 + 3 v_0 v_{-1}) + 2 v_2^3 v_1^3 v_0 + \nonumber \\
&& v_2^2 v_1^2 v_0 (2 v_1 v_0 +2 v_1 v_{-1} + 2 v_0 v_{-1} + v_{-1} v_{-2}) + v_2 v_1 v_0 v_{-1} (2 v_1 v_0 + v_{-2} v_{-3}) (v_1 + v_{-1})\,.
\end{eqnarray}

\end{document}